# Hydex Glass and Amorphous Silicon for Integrated Nonlinear Optical Signal Processing

R. Morandotti[2] and D. J. Moss[1,2]

[1] INRS-EMT, 1650 Boulevard Lionel Boulet, Varennes, Québec, Canada, J3X 1S2

[2] School of Electrical and Computer Engineering, RMIT University, Melbourne, Victoria, 3001 Australia

dmoss@physics.usyd.edu.au

*Abstract*

Photonic integrated circuits that exploit nonlinear optics in order to generate and process signals all-optically have achieved performance far superior to that possible electronically - particularly with respect to speed. Although silicon-on-insulator has been the leading platform for nonlinear optics for some time, its high two-photon absorption at telecommunications wavelengths poses a fundamental limitation. We review the recent achievements based in new CMOS-compatible platforms that are better suited than SOI for nonlinear optics, focusing on amorphous silicon and Hydex glass. We highlight their potential as well as the challenges to achieving practical solutions for many key applications. These material systems have opened up many new capabilities such as on-chip optical frequency comb generation and ultrafast optical pulse generation and measurement. **Keywords-component; CMOS Silicon photonics, Integrated optics, Integrated optics Nonlinear; Integrated optics materials**

## I. INTRODUCTION

Photonic integrated circuits that exploit nonlinear optics in order to achieve all-optical signal processing have been demonstrated in silicon-on-insulator (SOI) nanowires [1] and chalcogenide glass (ChG) waveguides [2, 3]. Some of the key functions that have been demonstration include all-optical logic [4], demultiplexing from 160Gb/s [5] to over 1Tb/s [6] down to much lower base-rates via four-wave mixing (FWM), to optical performance monitoring using slow light at speeds of up to 640Gb/s [7-9], all-optical regeneration [10,11], and many others. The 3$^{rd}$ order nonlinear efficiency of all-optical devices depends on the waveguide nonlinear parameter, $\gamma = \omega\, n_2 / c\, A_{eff}$ (where $A_{eff}$ is the waveguide effective area, $n_2$ the Kerr nonlinearity, and $\omega$ the pump frequency). It can also be increased by using resonant structures to enhance the local field intensity. High index materials, such as semiconductors and ChG, offer excellent optical confinement and high values of $n_2$, a powerful combination that has produced extremely high nonlinear parameters of $\gamma$= 200,000 W$^{-1}$ km$^{-1}$ for SOI nanowires [1], and 93,400 W$^{-1}$ km$^{-1}$ in ChG nanotapers [2].

However, silicon suffers from high nonlinear losses due to intrinsic two-photon absorption (TPA) and the resulting generated free carriers that tend to have very long lifetimes. Even if the free carriers could be eliminated by using p-i-n junctions, however, silicon's very poor intrinsic nonlinear figure of merit (FOM = $n_2$ / ($\beta\, \lambda$), where $\beta$ is the two-photon absorption coefficient) of around 0.3 in the telecom band is far too low to achieve extremely high performance all-optical processing. While ChG have a considerably higher nonlinear FOM, their fabrication processes are at a much earlier stage of development. The consequences of silicon's low FOM was clearly illustrated in breakthroughs at longer wavelengths in the SWIR band, where the TPA vanishes [12, 13]. While TPA can be used to advantage for all-optical functions [14-17], for the most part in the telecom band silicon's low FOM is a fundamental limitation, since it is a material property that is intrinsically dependent on the bandstructure and so it cannot be improved.

In 2008-2010, new platforms for nonlinear optics that are CMOS compatible were introduced, including Hydex and silicon nitride [17-26]. These platforms exhibit negligible nonlinear absorption in the telecom band, and have since been the basis of many key demonstrations. They have revolutionized micro-resonator optical frequency combs as well as and ultrashort modelocked lasers [31]. They show great promise for applications to ultrahigh bandwidth telecommunications [32]. Furthermore, they have shown great promise for applications to quantum optics [33].

We review recent progress made in Hydex glass [26] as well as amorphous silicon. Amorphous silicon is a relatively new and promising platform for nonlinear optics. The first integrated CMOS compatible integrated optical parametric oscillators were reported in 2010 [19, 34], showing that Kerr frequency comb sources could be realised in chip form by using ring resonators with relatively modest Q-factors compared to the extremely high Q's of micro-toroid structures [35]. Continuous wave "hyper-parametric" oscillation in a micro-ring resonator with a Q factor of 1.2 million was demonstrated with a differential slope efficiency 7.4% for a single oscillating mode out of a single port, a low CW threshold power of 50mW, and a variable range of frequency spacings from 200GHz to > 6THz.

Following this, we demonstrated a stable modelocked laser with pulse repetition rates from 200GHz to 800GHz [21]. We also demonstrated novel functions based on 45cm long spiral waveguides. This included a device capable of simultaneous measurement of both the amplitude and phase of ultrafast optical pulses. This device was based on Spectral Phase Interferometry by Direct Electric-field Reconstruction (SPIDER) [22]. These long spiral waveguides also achieved high optical parametric gain approaching +20dB [25].

The success of this platform arises from its very low linear loss as well as reasonably high nonlinearity parameter ($\gamma \cong 233W^{-1}km^{-1}$) and, most importantly, negligible nonlinear loss (TPA) up to intensities of $25GW/cm^2$ [20]. The low loss, design flexibility, and CMOS compatibility of these devices will go a long way to achieving many key applications such as multiple wavelength sources for telecommunications, computing, metrology and many other areas.

before over-coating with silica glass. Propagation losses were < 0.06dB/cm and pigtail coupling losses to standard fiber was $\cong$ 1.5dB / facet. The ring resonator had a FSR of 200GHz and a resonance linewidth of 1.3pm, with a Q of 1.2 million. The dispersion is anomalous [17] over most of the C-band with a zero dispersion point for TM polarization near ~1560nm with $\lambda$ < 1560nm being anomalous and $\lambda$ > 1560nm normal.

Figure 1 shows the OPO spectrum for a TM polarized pump at 1544.15nm at a power of 101mW. Oscillation initiates near 1596.98nm, 52.83nm away from the pump, with frequency spacing of 53nm. This agrees well with the calculated peak in the modulational instability (MI) gain curve [19] which occurs near ~1590nm. Figure 1 also shows the output power of the mode at 1596.98nm from the drop port, versus pump power, showing a differential slope efficiency of 7.4%. The maximum output power at 101mW pump at 1544.15nm was 9mW in all modes out of both ports, representing a total conversion efficiency of 9%. When pumping at 1565.19 nm in the normal dispersion regime, oscillation was not achieved. Pumping at zero dispersion (1558.65nm) resulted in a spacing of 28.15nm, which also agreed with the peak in the MI gain profile.

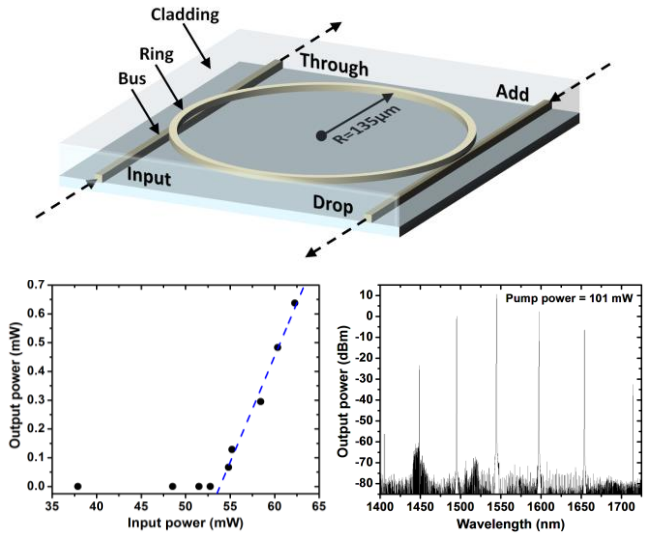

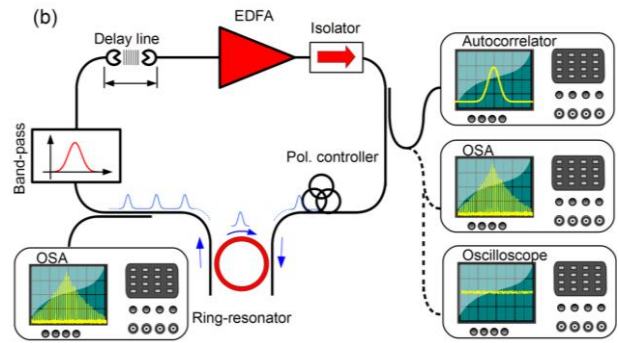

Figure 1. (top) Output spectra of hyperparametric oscillator near threshold (50mW) (top let) and at full pumping power (101mW, top right). Output spectrum when pumping closer to the zero dispersion point (bottom left) and output power of a single line, single port vs pump power.

Figure 2. Experimental configuration for filter-driven dissipative four wave mixing based modelocked laser.

## II. INTEGRATED COMB SOURCE

Figure 1 shows a ring resonator that was the basis of the integrated optical parametric oscillator. It is a four port micro-ring resonator with radius $\cong$ 135μm in waveguides havng a cross section of 1.45μm x 1.5μm, buried in $SiO_2$. The waveguide core is Hydex glass with n=1.7 and a core-cladding contrast of 17% [19]. The glass films were deposited by plasma enhanced chemical vapor deposition (PECVD) and subsequently processed by deep UV photolithography using stepper mask aligners followed by dry reactive ion etching,

Since these reports of integrated freqency comb sources [19, 34, 35] activity in this area has increased substantially [36-40] with demonstrations of cavity solitons [31], use of these combs in ultrahigh bandwidth systems [32], arbitrary waveform generation [40] and many others. The approach to generating ultrashort modelocked optical pulses using microresonators, that has achieved the greatest success in terms of stability to date has been based on filter driven four-wave mixing (FD-FWM) [21], and we turn to this next.

## III. SELF-LOCKED LASERS

Passively mode-locked lasers generate the shortest optical pulses [41 - 53]. Many approaches have been proposed to achieve very high and flexible repetition rates at frequencies well beyond active mode-locking, from very short laser cavities with large mode frequency spacings [21,41,45-48], where a very high repetition rate is achieved by simply reducing the pulse round-trip time, to lasers where multiple pulses circulate each round trip [33,47,48]. In 1997 [48] dissipative FWM [49, 51] was introduced, where a Fabry Pérot filter was inserted in the main cavity to suppress all but a few periodically spaced modes, leading to a train of pulses with a very high repetition rate. Although dissipative FWM yielded transform limited pulses at very high repetition rates, a fundamental problem is super-mode instability where multiple pulses circulate in a cavity. This is a consequence of the much smaller cavity mode frequency spacing of a few megahertz or less, which allows many modes to oscillate within the Fabry Pérot filter bandwidth, thus producing extremely unstable operation [52].

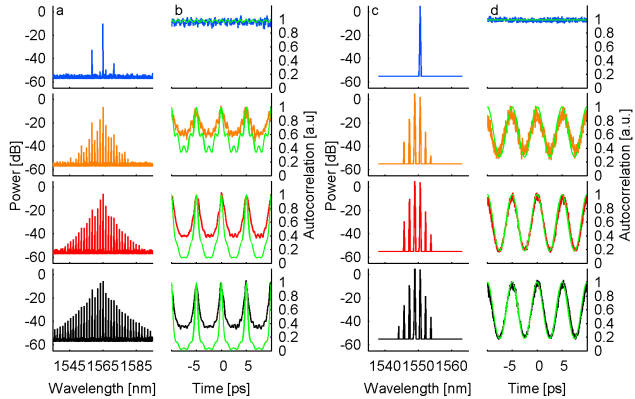

Figure 3. (left) Optical output for long cavity length (unstable) laser and (right0 output of short cavity (stable laser). Green curves are theoretical autocorrelation plots.

Figure 2 shows the configuration of the first mode-locked laser [21, 23] based on a nonlinear monolithic high-Q resonator that achieved extremely stable operation at high repetition rates while maintaining very narrow linewidths. The resonator is used as both filter and nonlinear element. This mode-locking scheme is termed filter-driven four-wave-mixing (FD-FWM). It operates in a way which is in stark contrast to traditional dissipative FWM schemes where the nonlinear interaction occurs in the fibre and is then filtered separately by a linear Fabry Pérot filter. The micro-ring resonator is embedded in an Erbium doped-fibre loop cavity containing a passband filter with a bandwidth large enough to pass all of the oscillating lines. A delay line controls the phase of the main cavity modes with respect to the ring modes.

Figure 3 compares the optical output of a laser based on a 33m long cavity along with one that used a very short EYDFA (Erbium Ytterbium), with a fiber loop length of only 3m. The two configurations had significantly different main cavity lengths (3m and 33m) with different FSRs (68.5MHz and 6MHz) as well as different saturation powers. Figure 3 also compares the optical spectra of the pulsed output along with the temporal traces obtained by an auto-correlator for the two systems at four pump powers. The pulses visible in the autocorrelation trains had a temporal duration that decreases noticeably as the input power increases, as expected for a typical passive mode-locking scheme.

From these plots it would appear that the long laser had better overall performance since its pulsewidth was shorter. However, the key issue of laser stability is better illustrated by a comparison between the experimental autocorrelation traces with the calculated traces (green) in Figure 3. While a perfect match is found for the short length EDFA case, the long cavity design shows a considerably higher background, thus clearly distinguishing unstable from stable laser operation. To quantify the pulse-to-pulse stability we recorded the electrical radio-frequency (RF) spectrum of the envelope signal, collected at the output using a fast photo-detector. Unstable oscillation was always observed [21] for the long cavity due to the presence of a large number of cavity modes oscillating in the ring resonance. In contrast, the short-cavity could easily be stabilized to give very stable operation by adjusting the main cavity length in order to center a single main cavity mode with respect to the ring resonance, thus eliminating any main cavity low-frequency beating. This self-locked approach has also been applied to CW operation where a two comb lines at each resonance were obtained to yield an ultra-pure radio frequency beat tone [23] and we turn to this next.

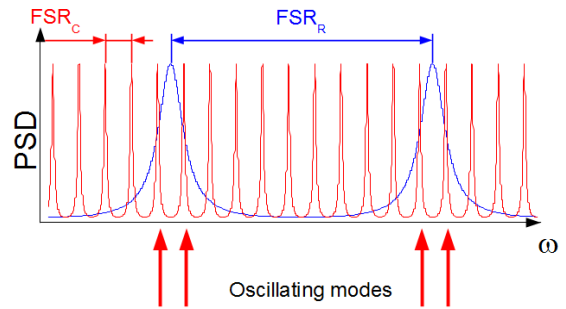

Figure 4. Mode configuraton for dual-comb operation. The main cavity modes are aligned using the free-space delay line so that two modes are placed symmetrically with respect to the ring resonator modes, thus allowing 2 – and only 2 - main cavity modes to oscillate simultaneously.

Figure 4 shows the mode configuration for the FD-FWM laser scheme shown in Figure 2, when operated in a novel stable operating regime where two main cavity modes within each micro-resonator linewidth are allowed to oscillate [23]. This operating regime leads to the formation of two spectral comb replicas separated by the free-spectral range (FSR) of the external main cavity which in this case is FSRMC= 65MHz. The beating of the two combs generates a sinusoidal modulation of the output pulse train at the radio-frequency of the main cavity FSR, and being in the RF regime is readily detectable with standard photodiodes and RF equipment. In addition to providing a high fidelity, ultra-stable RF tone, this approach yields key information about the phase noise and relative modal frequency coherence of this type of laser that is otherwise only accessible using very difficult and indirect optical measurement methods.

Figure 5 shows the results of this operating scheme. On the timescale of the optical pulses, this operating regime is similar to that of the stable case of Fig. 3. The autocorrelation here also reveals a stable pulsation consistent with the calculated autocorrelation trace obtained by assuming that the lines in the optical spectrum are in phase. However, the temporal RF signal measured with a high speed photodiode (Fig. 5(a)) along with its RF spectrum (Fig.5 (b,c)), shows highly coherent beating at 65MHz, indicating that two distinct modes of the main cavity are oscillating for each micro-ring resonator line. Using a filter we verified that the beating originated from mode doublets occurring in every excited resonance of the micro-ring.

These measurements indicate that the system was oscillating with two main cavity modes distributed around a ring resonance peak. The laser oscillated in a stable mode-locking manner with an output spectrum consisting of a number of closely spaced main cavity mode doublets. An extinction ratio of the beating (Fig. 5 (a)) that exceeds 80% demonstrates a good balance in energy of the two comb replica. The secondary peak around 131 MHz in Fig.5 (b) is a second harmonic of the RF beat frequency, but is at a much lower power level (-23dB down with respect to the main RF peak). Fig. 5(c) shows the particularly narrow linewidth of the RF beat frequency estimated to be < 10kHz (FWHM).

This beating is related to the frequency difference between the two oscillating comb replicas and so its spectrum is an accurate reflection of the dynamics of the line-to-line spacing of the main cavity modes. This confirms the instantaneous frequency locking between the main cavity oscillating lines.

These results demonstrate the versatility of the FD-FWM scheme as a function of the main cavity length. By adjusting the relative phase between the main cavity modes and the ring resonator, not only is it possible to eliminate all the low frequency instabilities arising from super-mode beating and EDFA gain switching in order to obtain stable pulsed operation, but it is also possible to achieve stable oscillation of a dual comb separated by the FSR of the main cavity.

An important aspect of these results is that the generated RF modulation is coherent with the emission of a 200GHz optical pulse train because it arises from beating between cavity modes. This is qualitatively different from modulating the pulse train with an external modulator, for example. To first order the modulation can be thought of as a Phased Locked sub-carrier since its frequency shifts with the 200GHz rep-rate. Although it is not necessarily an integer fraction of the repetition rate, obtaining a locked sub-carrier requires one to "read" the pulse train with a Phase Locked Loop, but in this case the repetition rate (200GHz) is well beyond the capability of typical detectors. Instead, in this case the signal already carries the information that is useful, for example, to synchronize slower sources for time-division multiplexing applications. Moreover, the measurement of the amplitude noise of this configuration gives important information on the phase stability of the system, as the line width of the AC peak at 65.8MHz in Fig. 5(c) is a direct measurement of the stability of the FSR of the main cavity, which is better than 0.01% for the results presented here. This measurement reflects the stability of the ultrafast comb, and in principle can be employed to stabilize the whole system at a frequency (65.8 MHz) that can be easily followed by electronics.

IV. RADIO FREQUENCY SPECTRUM ANALYZER

Photonics offers the capability of generating and measuring ultrashort optical pulses, with bandwidths of many THz and at repetition rates of hundreds of GHz. Performing temporal diagnostics at these speeds is extremely difficult and yet essential to achieve high optical signal fidelity of fundamental noise parameters such as time jitter and amplitude noise, critical for achieving the maximum performance of many devices such as high frequency - clock optical modules [54, 55]. The traditional way of measuring the RF spectrum consists of recording the temporal intensity profile by an ultra-fast photo detector and then processing this signal, but this approach is limited to around 50 GHz.

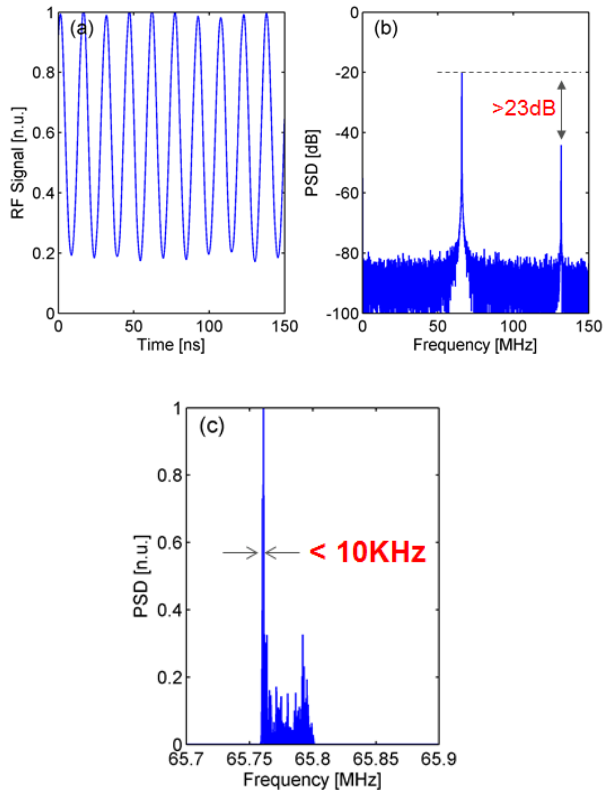

Figure 5. Dual-comb operation. RF signal of the laser output in time (a) and spectrum (b), showing a zoom-in of the RF spectrum with a monochromatic beating at 65.8 MHz. Figure (c) shows a high resolution plot of the AC component at 65.8MHz, showing a linewidth narrower than 10kHz.

A key breakthrough was the all-optical RF-spectrum analyzer, introduced by Dorrer and Maywar [56], based on optical mixing between a signal and CW probe via the Kerr ($n_2$) nonlinearity. In this approach, a single measurement of the CW probe optical spectrum with an optical spectrum analyzer (OSA) yields the intensity power spectrum of the signal under test. This approach can achieve much broader bandwidths than electronic methods, with a trade-off between sensitivity and bandwidth, or between the nonlinear response and total dispersion of the waveguide. Since then much progress has been made in realizing this device in integrated form [57 - 61]. Increasing the device length enhances the nonlinear response but results in increased dispersion that reduces the frequency response. For this reason, an optical integration platform with high nonlinearity and low net dispersion (waveguide plus material) is highly desirable. The first demonstration of an integrated all-optical RF spectrum analyzer [57] was achieved in chalcogenide waveguides in only a few centimeters of length. This was followed by a device on a silicon nanowire [59] and was subsequently used to monitor dispersion of ultrahigh bandwidth coherent signals [61].

Recently [60], we reported an integrated RF spectrum analyzer based on Hydex glass [26]. Figure 6 shows the device configuration of the RF spectrum analyzer while Figure 7 shows its measured frequency response showing a 3dB bandwidth of about 2.6THz, limited by our system measurement capability. We believe the intrinsic bandwidth is substantially higher than this since the simple theoretical prediction used in [52, 53] yields a bandwidth more than 100THz. In practice this would likely be limited by higher order dispersion, mode cutoff and even absorption bands, since the simple model of [56, 57] does not include these effects. We used this device to measure the RF frequency response of the ultrahigh repetition rate laser discussed in the last section that emitted sub picosecond pulses at repetition rates of 200 GHz and 400GHz. This device allowed us to analyze these lasers according to the noise burst model, which identifies very rapid intensity fluctuations of the laser pulses as the main source of noise [62, 63]. The RF spectra for both lasers show sensitivity to high frequency noise not detectable by other methods.

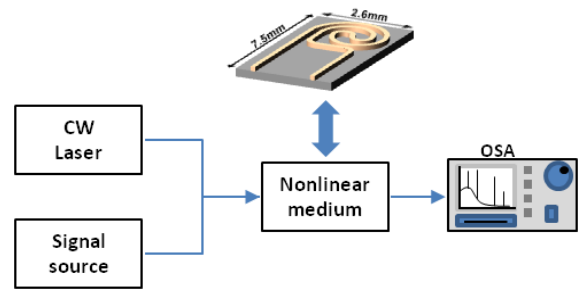

Figure 6. Experimental configuration for filter-driven dissipative four wave mixing based modelocked laser.

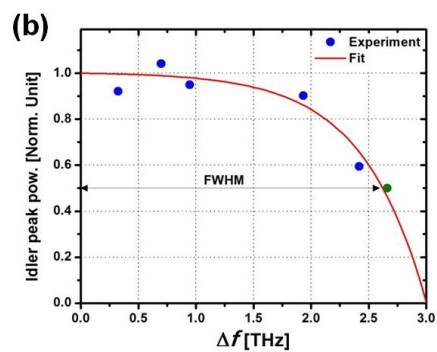

Figure 7. Experimental configuration for filter-driven dissipative four wave mixing based modelocked laser.

## V. AMORPHOUS SILICON

Amorphous silicon has been studied as a nonlinear material for some time [64], but it has only been recently that it was proposed [65] as a CMOS compatible alternative to SOI for nonlinear optics. It has been shown to have a much larger FOM than SOI [66- 73], with reports of a FOM ranging from near unity [68, 69] to as high as 2 near 1550nm [70]. This was the basis of reports of high optical parametric gain of > +26dB in the C-band [69]. Table I compares the nonlinear properties of Hydex, silicon nitride, and crystalline and amorphous silicon where it can be seen that a-Si shows a significant improvement in both nonlinearity and FOM over SOI. Recently, we demonstrated [71] a significant enhancement in stability of amorphous silicon [69, 70], along with a record high FOM of 5, which is more than ten times that of SOI, as well as a $\gamma$ factor of almost five times that of silicon. This may seem surprising since Kramers - Kronig relations normally imply that increasing the bandgap to decrease nonlinear absorption decreases the nonlinear response. In silicon, however, the real part of the nonlinear susceptibility is dominated by direct transitions [74 - 78] whereas TPA in the telecom band arises from indirect transitions. For a-Si therefore, it could be that an increase in the *indirect* bandgap (reducing TPA) could be accompanied by a decrease in the *direct* bandgap (increasing the Kerr nonlinearity. The direct bandgap in silicon, which is around 3.5eV [79], is largely responsible for determining the real part of the nonlinear response whereas the indirect bandgap dominates transitions involving multiphoton absorption. Therefore decreasing the direct bandgap will not significantly increase the two-photon absorption and conversely, increasing the indirect bandgap will not significantly affect $n_2$.

Finally, a key goal for all-optical chips is to reduce device footprint and operating power, and the dramatic improvement in the FOM of a-Si raises the possibility of using slow-light structures [7 - 9] to allow devices to operate at mW power levels with sub-millimeter lengths.

## VI. CONCLUSION

We review the progress made in CMOS compatible chips for nonlinear optics and optical signal processing based on the Hydex glass platform and amorphous silicon. These devices have significant potential for applications requiring CMOS compatibility for both telecommunications and on-chip WDM optical interconnects for computing and many other applications.

## Table I
Nonlinear parameters for key materials

|  | a-Si | c-Si | SiN | Hydex |
|---|---|---|---|---|
| $n_2$ (x fused silica[1]) | 700 | 175 | 10 | 5 |
| $\gamma$ [W$^{-1}$m$^{-1}$] | 1200 | 300 | 1.4 | 0.25 |
| $\beta_{TPA}$ [cm/GW] | 0.25 | 0.9 | 0 [2] | 0 [3] |
| FOM | 5 | 0.3 | ∞ | ∞ |

[1] $n_2$ for fused silica = 2.6 x 10$^{-20}$ m$^2$/W
[2] no nonlinear absorption has been observed in SiN nanowires.
[2] no nonlinear absorption has been observed in Hydex waveguides up to intensities of 25GW/cm$^2$.